\documentclass[12pt]{iopart}

\usepackage{color}
\usepackage{graphicx}
\usepackage{amssymb}

\usepackage{hyperref} 

\begin{document}
 {\small \today}
\ \\ \ \\ 
\title[Results and conjectures on a toy model of depinning]
 {Results and conjectures on a toy model of depinning}

\author{
 Bernard Derrida$^{1}$ and  Zhan Shi$^2$,}

\address{$^1$ Coll\`ege de France, Universit\'e PSL, 11 place Marcelin Berthelot, F-75231 Paris Cedex 05, France}
\address{$^2$  Laboratoire de Probabilit\'es, Statistique  et  Mod\'elisation  (LPSM), Sorbonne Universit\'e, 4 place Jussieu, F-75252 Paris Cedex 05, France, }
\ead{bernard.derrida@phys.ens.fr,\\ zhan.shi@upmc.fr }

\begin{abstract}
We review recent results and conjectures  for a simplified version
  of  the depinning problem in presence  of disorder
 which was introduced by Derrida and Retaux in 2014.
For  this toy model, the depinning transition has been predicted 
to be of the    Berezinskii--Kosterlitz--Thouless type.  Here we discuss under which integrability conditions this prediction can be proved and how it is modified otherwise.
\end{abstract}

\ \\ \ \\ \ \\ 
\begin{center}
{\it Submitted to  
the Moscow Mathematical Journal 
\\ for  the volume dedicated to the memory of
Roland Dobrushin} 
\end{center}
\ \\ \ \\
\maketitle
\normalsize

\def\log{\ln}

\section{Introduction}
 The depinning transition in one dimension in presence of disorder has been for many years \cite{luck,derrida-hakim-vannimenus,giacomin}
a challenging  problem in the theory of disordered systems (see the appendix). 
Trying to determine the critical behavior of the free energy at this transition
 has led to introduce several simplified versions of the problem. The goal of the  present paper is to review the main recent results and main open questions for  one of these simplified versions (see equation (\ref{recur}) below).

Let us first formulate a general question posed by the version of the depinning problem on a hierarchical lattice.
Given a non-negative random variable $X_0$, what can be said~\cite{derrida-hakim-vannimenus} of the law of the random variable $X_n$ defined recursively via the following deterministic formula: for all $n\ge 0$,
\begin{equation}
X_{n+1} = G(X_n^{(1)} + X_n^{(2)}),
\label{system}
\end{equation}
where $X_n^{(1)}$ and $X_n^{(2)}$ are independent random variables having the law of $X_n$? Here, we assume that   $G: \, [0, \, \infty) \to [0, \, \infty)$ is a continuous, piecewise differentiable,  convex,   non-decreasing function  such that $G(0)=0$,\    $0 \le G'(0) < {1 \over 2}$ and that 
\begin{equation}
G(x) = x-c + o(1), \qquad x\to \infty,
	\label{function}
	\end{equation}
	for some constant $c>0$.

	 If $X^*$ is the non-zero positive solution of   $X^* = G(2 X^*)$,  it is clear that the free-energy ${\cal F}_\infty$  defined by
		\begin{equation}
		{\cal F}_\infty \equiv \lim_{n \to \infty} \frac{\langle X_n\rangle}{2^n} \, ,
			\label{free-energy-bis}
		\end{equation}
		vanishes if the support of $X_0$ is included  in the interval $[0,X^*]$ while it is strictly positive when this support is included in $(X^*,\infty)$. Therefore  as one varies the initial distribution of $X_0$, one should cross a critical manifold (in the space of  these distributions)  which separates the domain where  ${\cal F}_\infty=0$  from the domain where ${\cal F}_\infty>0$.

Given $G$, the basic questions we ask ourselves are: How to characterize this
critical manifold (and to start with, whether it exists)? 
 How does  ${\cal F}_\infty$ vanish as one approaches this manifold? 

As explained in the appendix an  example   of such a $G$, motivated by the depinning problem on a hierarchical lattice, is 
	\begin{equation}
	G(x) = \log \Big( \frac{e^x + b-1}{b} \Big), \qquad x \ge 0,
		\label{G}
	\end{equation}
	where $b > 2$ is a fixed constant (in this case $X^*=\log(b-1)$). 

	A simplified  recursion (which can be obtained,  up to a rescaling on $x$,  as  the $b \to \infty$ limit of (\ref{G}))   was proposed in \cite{derrida-retaux}:
\begin{equation}
		X_n= \max\Big[ X_{n-1}^{(1) }+ X_{n-1}^{(2)} -1, \, 0\Big] . 
			\label{recur}
		\end{equation}
To avoid discussing an obvious case we will  consider everywhere  that $P(X_0 \ge 2) >0$.
		The free energy (\ref{free-energy-bis})  is then defined as follows:
			\begin{equation}
		{\cal F}_\infty \equiv \lim_{n \to \infty} \frac{\langle X_n\rangle}{2^n} \, .
		\label{free-energy}
		\end{equation}
		The  existence of the limit is  immediate since $n\mapsto \frac{\langle X_n \rangle}{2^n}$ is non-increasing.  Here  we wish  to understand how the free-energy (\ref{free-energy}) depends on the law of $X_0$ in particular near the transition between the phase where ${\cal F}_\infty=0$ and the  phase where ${\cal F}_\infty >0$.

		When $X_0$ is {\it integer-valued} and non-negative (which is the only case we consider in the present review), it is more convenient to study (\ref{recur})  via the moment generating function:
		$$
			H_n(z) \equiv \langle z^{X_n} \rangle .
			$$
			Then (\ref{recur}) is equivalent to the following recursion
			\begin{equation}
		H_{n+1}(z) = \frac{H_{n}(z)^2 - H_n(0)^2}{z} + H_n(0)^2 \, .
			\label{rec1}
		\end{equation}

		 The  simplest approach 
			to study  (\ref{rec1}) would be to look for its fixed points. This turns out to be fruitless because all fixed points  have negative coefficients in their expansion in powers of $z$ and therefore cannot be the generating functions of probability distributions.

				The recursion (\ref{rec1}) was first studied by Collet, Eckmann, Glaser and Martin 
				\cite{collet-eckmann-glaser-martin,collet-eckmann-glaser-martin2} in the context of  spins glasses. A remarkable result in \cite{collet-eckmann-glaser-martin} is the characterization of the critical manifold: defining
				\begin{equation}
		\Delta \equiv 2 H'_0(2)  - H_0(2),
			\label{Delta-def}
		\end{equation}
		(with the convention $\Delta \equiv \infty$ if $\langle X_0 2^{X_0} \rangle =\infty$), then they proved that 
		\begin{eqnarray*}
		&&{\cal F}_\infty = 0 \qquad \hbox{if } \Delta \le 0, \\
			&&{\cal F}_\infty > 0 \qquad \hbox{if } \Delta >0. 
			\end{eqnarray*}
		As such, the critical manifold is 
			\begin{equation}
		\Delta=0,
			\label{manifold}
		\end{equation}
		and the value of $\Delta$ measures the distance to the critical manifold.


		Our goal is to give an overview on various predictions and rigorous results about the recursive equation (\ref{recur}). Most of these predictions and results concern systems on  the critical manifold (\ref{manifold}), with the  exception of Sections \ref{s:energy} and \ref{s:heavytail} where we discuss the free energy for slightly supercritical systems.
In the following   we will  always consider that the random variable 
\begin{equation}
 X_0 \textrm{  is integer-valued and non-negative}. 
\label{cond1}
\end{equation}
In most sections we will also impose either the integrability condition 
\begin{equation}
\langle X_0 \, 2^{ X_0} \rangle < \infty
\label{cond2}
\end{equation}
which is always satisfied on the critical manifold (\ref{Delta-def},\ref{manifold})
or the stronger condition 
\begin{equation}
\langle X_0^3 \, 2^{ X_0} \rangle < \infty \ . 
\label{cond3}
\end{equation}

			The paper \cite{HMP} presents an interesting continuous-time related model. When the initial distribution is exponential, it answers all the analogous questions for the continuous-time model, sometimes with different numerical values of constants due to the continuous-time setting.  


		\section{The sustainability probability $P(X_n >0)$}
		\label{s:sustainability}
		In this section we discuss the large $n$ decay  of $P(X_n>0)$ when the distribution of $X_0$ is on the critical manifold  (see (\ref{Delta-def},\ref{manifold}))  implying that (\ref{cond2}) is satisfied.

			It has been shown  \cite{collet-eckmann-glaser-martin,bmxyz_questions}   that $P(X_n >0) \to 0$ on the critical manifold. The question is about the rate at which this probability goes to 0. 

			\subsection{Conjectures}

		It was first predicted in both \cite{collet-eckmann-glaser-martin} and \cite{derrida-retaux}, but without precision on integrability condition on $X_0$, that 
			\begin{equation}
		P(X_n>0) \sim \frac{4}{n^2} , \qquad n\to \infty.
			\label{factor4}
		\end{equation}
		[Notation: $a_n \sim b_n$ means $\lim_{n\to \infty} \frac{a_n}{b_n} =1$.] 

			In \cite{6authors2}, it was realized that (\ref{factor4}) should only be valid when  condition (\ref{cond3}) is satisfied ($\langle X_0^3 \, 2^{X_0}\rangle <\infty$). For  $P(X_0=k) \simeq c_0 k^{-\alpha}2^{-k}$, $k\to \infty$, for some $\alpha \in (2, \, 4]$ (implying that $\langle X_0^3 \, 2^{X_0}\rangle =\infty$), the equation (\ref{factor4}) should be replaced by
\begin{equation}
P(X_n>0) \sim \frac{c(\alpha)}{n^2} , \qquad n\to \infty ,
\label{factor_alpha}
\end{equation}
\noindent where $c(\alpha) := \frac{\alpha(\alpha-2)}{2}$. 

\subsection{Results}

It has been  proved in \cite{6authors2} that if 
\begin{equation}
\langle z^{X_0}\rangle <\infty \; \hbox{ for some } z>2,
\label{assump:z>2}
\end{equation}
then 
\begin{equation}
P(X_n >0) = \frac{1}{n^{2+o(1)}}, \qquad \langle X_n\rangle = \frac{1}{n^{2+o(1)}}, \qquad n\to \infty.
\label{P(Xn>0)+E(Xn)}
\end{equation}
Without the assumption (\ref{assump:z>2}), much less in known \cite{bmxyz_questions}: 
\begin{equation}
P(X_n >0) = O\left(\frac{1}{n}\right), \qquad \sum_n P(X_n >0) <\infty.
\label{O(1)}
\end{equation}
The first inequality in (\ref{O(1)}) follows from the forthcoming (\ref{H(2)-1}) and the Markov inequality $P(X_n>0) \le \langle 2^{X_n}\rangle -1$.

Note  that without the assumption (\ref{assump:z>2}) it has not even  been proved that $O\left(\frac{1}{n}\right) $ in (\ref{O(1)}) can be replaced  by $o\left(\frac{1}{n}\right)$.

\section{Weak convergence}
\label{s:weak_convergence}

In this section we  assume again that the system is on the critical manifold (\ref{Delta-def},\ref{manifold}). Being on the critical manifold implies $X_n \to 0$ in $L^1$ \cite{bmxyz_questions}. However, since $P(X_n >0)>0$, we are entitled to condition on the event $X_n>0$. The basic question is: given $X_n>0$, does $X_n$ converge weakly?

\subsection{Conjectures}

It is predicted in \cite{bmxyz_questions} without precision on integrability condition on $X_0$ that for any integer $k\ge 1$, we would have
\begin{equation}
\lim_{n\to \infty} P(X_n=k \, | \, X_n>0) = \frac{1}{2^k}\, .
\label{weak_cvg}
\end{equation}
In words, conditionally on being positive, $X_n$ would converge weakly to a geometric distribution of parameter $\frac12$.
 This  together  with (\ref{factor4}) would   imply 
\begin{equation}
\langle X_n \rangle \sim \frac{8}{n^2} \, 
\label{av1}
\end{equation}
when 
$\langle X_0^3 \, 2^{X_0}\rangle <\infty$.
If true, this would considerably refine the second part of (\ref{P(Xn>0)+E(Xn)}). 

On the other hand (see (\ref{factor_alpha}) and (\ref{weak_cvg})) for
  $P(X_0=k) \simeq c_0 k^{-\alpha}2^{-k}$ as  $k\to \infty$, for some $\alpha \in (2, \, 4]$, one expects (\ref{av1}) to be replaced by:
\begin{equation}
\langle X_n \rangle \sim \frac{ 2 c(\alpha)}{n^2} \, . 
\label{av2}
\end{equation}
\subsection{Results}

No rigorous result has yet been proved so far concerning weak convergence on the event $X_n>0$. It is not even clear how to prove (conditional) tightness: under suitable integrability condition on $X_0$, is it true that
$$
\lim_{k\to \infty} \limsup_{n\to \infty} P(X_n \ge k \, | \, X_n>0) = 0 ?
$$

\section{Scaling function in the generic case (i.e. when  $\langle X_0^3 \, 2^{X_0}\rangle <\infty$)}
\label{s:scale}
In this section we discuss the scaling form of $P(X_n=k)$ when the system is on the critical manifold (\ref{Delta-def},\ref{manifold}) and when the stronger integrability condition (\ref{cond3}) is satisfied ($\langle X_0^3 \, 2^{X_0}\rangle <\infty$).

\subsection{Conjectures}

We have seen in (\ref{weak_cvg}) that conditionally on  $X_n>0$, $X_n$ is expected to be geometric. It was even  predicted in  \cite{derrida-retaux}, without precision on integrability condition on $X_0$, that:
\begin{equation}
P(X_n =k ) = \frac{4}{n^2} \, \frac{1}{2^k} \, e^{-2k/n} + o\left(\frac{1}{n^2}\right), \qquad n\to \infty,
\label{P(Xn=k)}
\end{equation}
uniformly for integers $k\ge 1$. This would imply (\ref{factor4}) and (\ref{weak_cvg}).

We now believe (see Section \ref{s:stable}) that (\ref{P(Xn=k)}) would  only hold under the assumption $\langle X_0^3 \, 2^{X_0}\rangle <\infty$.

\subsection{Results}

The (conditional) exponential scaling function in (\ref{P(Xn=k)}) is not valid without the assumption $\langle X_0^3 \, 2^{X_0}\rangle <\infty$, and should be replaced by a more complicated scaling function  (see Section \ref{s:stable}).   

\section{Moment generating function}
\label{s:generating_fct}

Here again we assume that the system is on the critical manifold (\ref{Delta-def},\ref{manifold}) implying that condition (\ref{cond2}) is satisfied.

\subsection{Conjectures}

It was predicted in \cite{collet-eckmann-glaser-martin,bmxyz_questions} that (upon suitable integrability conditions on $X_0$),
\begin{equation}
\langle 2^{X_n}\rangle -1 \sim \frac{2}{n}\, ,
\label{H(2)-1:conj}
\end{equation}
and that for $z\in (0, \, 2) \backslash \{ 1\}$, 
\begin{equation}
\langle z^{X_n}\rangle -1 \sim \frac{z-1}{2-z} \, \frac{8}{n^2}\, .
\end{equation}

Also, we believe that upon suitable integrability conditions on $X_0$, for any integer $q\ge 2$, there  exist  constants $a(q)\in (0, \, \infty)$, depending on the law of $X_0$, such that
\begin{equation}
\langle X_n^{q} \, 2^{X_n}\rangle \sim a(q)\, n^{q-1} \, .
\end{equation}
(According to 
(\ref{P(Xn=k)}) 
one can even predict that  $a(q)= 2^{1-q} \, q!$ .) 
Note that for $q=0$, we have $\langle 2^{X_n}\rangle \to 1$ (see (\ref{H(2)}) below), and that for $q=1$, it follows from (\ref{manifold}) that $\langle X_n \, 2^{X_n}\rangle = \langle 2^{X_n}\rangle \to 1$.

\subsection{Results}

It has been shown   \cite{bmxyz_questions}  that 
\begin{equation}
\langle 2^{X_n}\rangle \to 1.
\label{H(2)}
\end{equation}
Assuming $\langle X_0^3 \, 2^{X_0}\rangle < \infty$, it was proved \cite{bmxyz_questions} that there exist constants $0<c_1 \le c_2<\infty$, depending on the law of $X_0$, such that for all $n\ge 1$,
\begin{equation}
\frac{c_1}{n} \le \langle 2^{X_n}\rangle - 1 \le \frac{c_2}{n} ,
\label{H(2)-1}
\end{equation}
and also that 
\begin{equation}
c_3\, n^2 \le \prod_{i=0}^{n-1}\langle 2^{X_i}\rangle \le c_4 \, n^2\, ,
\label{Pi}
\end{equation}
for some constants $0<c_3 \le c_4<\infty$ depending on the law of $X_0$. [The assumption $\langle X_0^3 \, 2^{X_0}\rangle < \infty$ is not needed for the second inequality in (\ref{Pi}).] Note that (\ref{Pi}) is in agreement with (\ref{H(2)-1:conj}).

For $\langle X_n^k \, 2^{X_n}\rangle$, it has been proved \cite{6authors2}, using the recursion (\ref{rec1}), that for any integer $k\ge 2$, if $\langle X_0^k\, 2^{X_0}\rangle <\infty$, then there exist constants $0<c_5(k) \le c_6(k) <\infty$ depending on the law of $X_0$, such that
\begin{equation}
c_5(k) \, n^{k-1} \le \langle X_n^k \, 2^{X_n}\rangle \le c_6(k) \, n^{k-1}, \qquad n\ge 1 \, .
\end{equation}
Under the stronger assumption that $\langle z^{X_0}\rangle<\infty$ for some $z>2$, the following is true~\cite{6authors2}: for any $\lambda>0$ and any integer $k\ge 1$, 
\begin{equation}
\liminf_{n\to \infty} \frac{\langle X_n^k \, 2^{X_n}\, {\bf 1}_{\{ X_n \ge \lambda n\} } \rangle}{n^{k-1}} >0 \, .
\end{equation}

\section{Free energy    (see (\ref{free-energy}))}
\label{s:energy}
In this section we discuss the critical behavior of the free energy (\ref{free-energy})
when condition (\ref{cond2}) is satisfied (i.e. when $\langle X_0 \, 2^{X_0}\rangle <\infty$).

{ When the system is supercritical, i.e. when  ${\cal F}_\infty>0$, there is no non-trivial example for which the value of ${\cal F}_\infty$ can be computed exactly. However, when a supercritical system is ``nearly supercritical", in the sense that $\Delta$, defined in (\ref{Delta-def}) and standing for the distance to the critical manifold, is positive but very small, one expects to see  universal behaviors of   the free energy. 

\subsection{Conjectures}

The model is expected to have a Berezinskii--Kosterlitz--Thouless type phase transition of infinite order; more precisely, the following prediction was made in \cite{derrida-retaux} (often referred to as the Derrida--Retaux conjecture for the free energy of the system): there would exist a constant $c_7 \in (0, \, \infty)$ depending on the law of $X_0$, such that when $\Delta\to 0+$,
\begin{equation}
{\cal F}_\infty = \exp\Big( - \frac{c_7+o(1)}{\Delta^{1/2}} \Big).
\label{conj_DR}
\end{equation}
As for  (\ref{factor4}) this prediction was made without precision on integrability. Now we believe that it   should only  be valid  when 
condition (\ref{cond3}) is satisfied ($\langle X_0^3\, 2^{X_0} \rangle <\infty$). Otherwise the exponent $1/2$ should be replaced by an exponent $\theta $ as in the result (\ref{stable}) below.

\subsection{Results}

The exponent $\frac12$ predicted in (\ref{conj_DR}) was proved under suitable integrability conditions on $X_0$. More precisely, it was established in \cite{6authors} that if $\langle X_0^3\, 2^{X_0} \rangle <\infty$, then
\begin{equation}
{\cal F}_\infty = \exp\Big( - \frac{1}{\Delta^{(1/2)+o(1)}} \Big), \qquad \Delta \to 0+. \label{eqA}
\end{equation}
Furthermore, the condition $\langle X_0^3\, 2^{X_0} \rangle <\infty$ was proved \cite{6authors} to be necessary for the validity of the exponent $\frac12$ in (\ref{conj_DR}): in particular if $P(X_0=k) \sim c_0 \, 2^{-k} k^{-\alpha}$ for $k\to \infty$, where $\alpha\in (2, \, 4]$ and $c_0 \in (0, \, \infty)$ is a constant, then  (\ref{eqA}) is replaced by 
\begin{equation}
{\cal F}_\infty = \exp\Big( - \frac{1}{\Delta^{\theta +o(1)}} \Big), \qquad \Delta \to 0+,
\label{stable}
\end{equation}
with $\theta = \theta(\alpha) := \frac{1}{\alpha-2}$. As such, (\ref{stable}) suggests another family of universal behaviors of the system under weaker integrability conditions i.e. when  $\langle X_0^3\, 2^{X_0} \rangle =\infty$. More discussions are made on this new classes of universality  in Section \ref{s:stable}.

\section{Open subtree}
\label{s:open_tree}
In this section we show that one can associate to each realization of $X_n$ a tree. We will assume that
the system is on the critical manifold (\ref{Delta-def},\ref{manifold}) so that  condition (\ref{cond2}) is satisfied.

There is a natural hierarchical representation for the system~\cite{derrida-retaux,collet-eckmann-glaser-martin} as a tree. Each vertex in the initial generation is attached with a spatial random variable; these random variables are i.i.d.\ having the distribution of $X_0$. More generally, for any $i \ge 0$, the random variables associated to each of the vertices at generation $i$ are i.i.d.\ having the same distribution as $X_i$. We are interested in the system leading to a given vertex at generation $n$. See Figure \ref{tree1}.

\begin{figure}[h]
\centerline{\includegraphics[width=7.5cm]{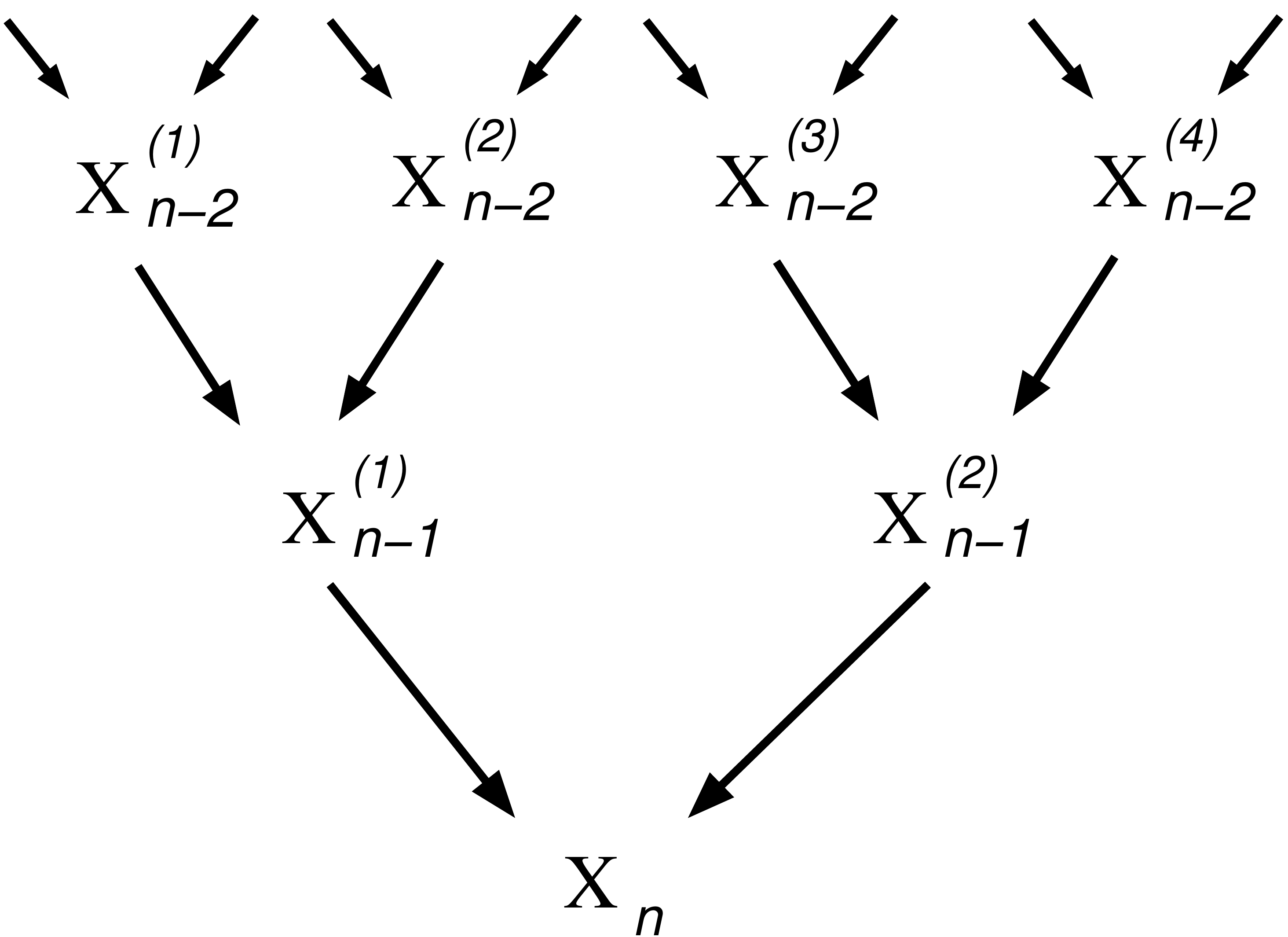}}
\caption{\small The random variable $X_n$ is a deterministic function (\ref{recur}) of $2^n$ independent realizations of the random variable $X_0$ located at the top  of a binary tree. }
\label{tree1}
\end{figure}


Each of the $2^n$ vertices at the initial generation has a unique path to the vertex at the bottom  of the tree (the one at generation $n$). A path is called {\it open} if, everywhere along the path, $a+b \ge 1$ in the  transformation $(a, \, b) \mapsto (a+b-1)^+$. So everywhere along an open path, one has $(a+b-1)^+=(a+b-1)$   (see Figure \ref{tree2}). 
\begin{figure}[h]
\centerline{\includegraphics[width=10cm]{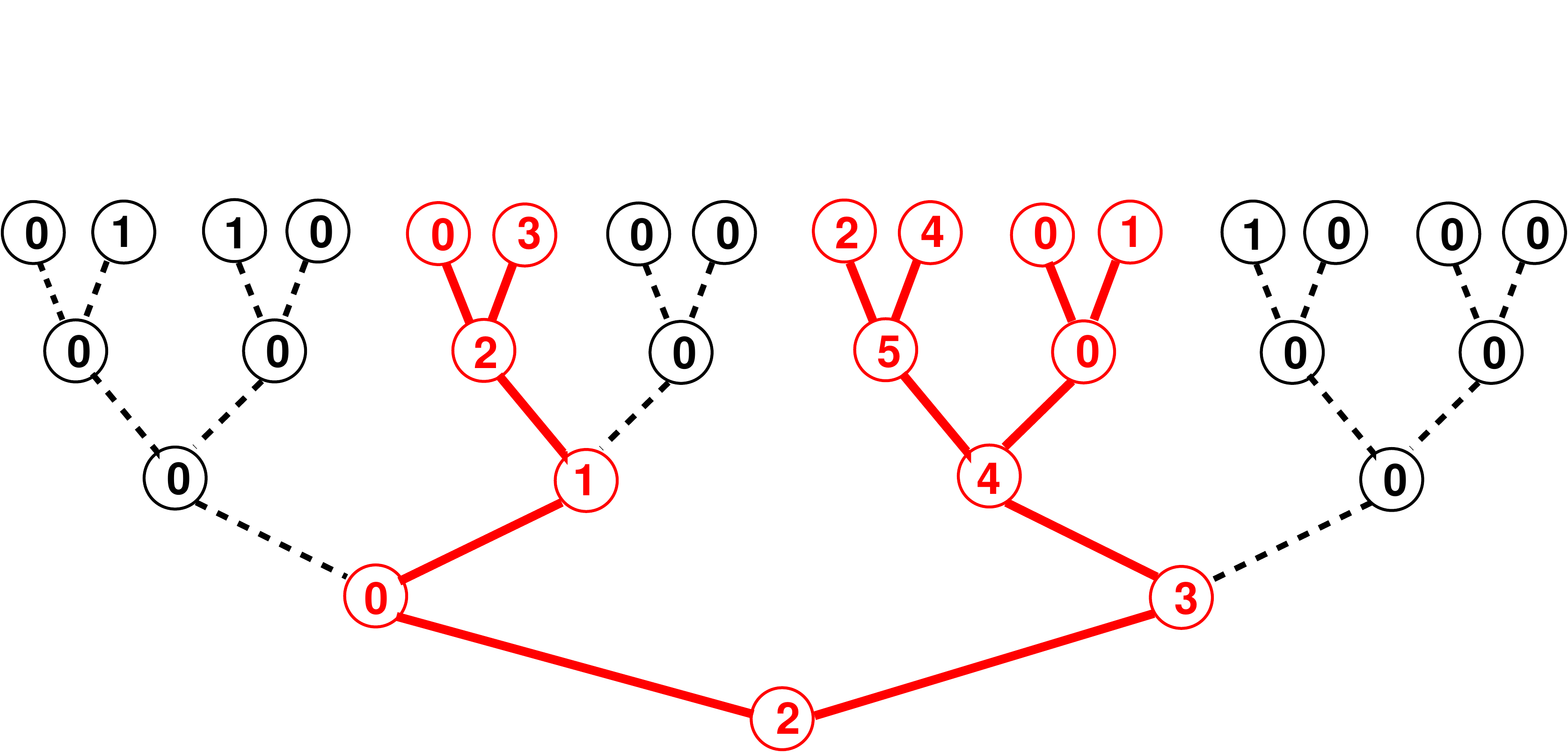}}
\caption{\small An example with $n=4$. Open paths are  represented by thick lines; they form the open subtree.  The other paths are represented by dashed lines. The number of open branches is $N_n=6$. Also, $N_n^{(0)} =2$, $N_n^{(1)} =1$, $N_n^{(2)} =1$, $N_n^{(3)} =1$, $N_n^{(4)} =1$, and $N_n^{(k)} =0$ for $k\ge 5$.}
\label{tree2}
\end{figure}


The set of all open paths forms a subtree. This subtree with $n$ generations,  denoted by $\mathbb{T}^{\mathrm{open}}_n$, is called the open subtree. We want to consider 
 quantitative characteristics of this subtree
$\mathbb{T}^{\mathrm{open}}_n$,
 such as  its   number of leaves $N_n$, or the joint distribution of $N_n$ and $X_n$. Note that it is possible to have $N_n \ge 1$ and $X_n=0$ simultaneously. 

For $k\in \{0, \, 1, \, 2, \ldots\}$, it is sometimes convenient to count $N_n^{(k)}$, the number of open paths starting from vertices  at the initial generation whose associated spatial value is $k$. (Note that it is possible to have $N_n^{(0)} \ge 1$). Obviously,
$$
N_n = \sum_{k=0}^\infty N_n^{(k)}\, .
$$
For $n=0$, we define $N_0^{(k)} \equiv {\bf 1}_{\{ X_0=k\} }$, and $N_0\equiv 1$. See Figure \ref{tree2}.

\subsection{Conjectures}

Let $x>0$. Upon suitable integrability conditions on $X_0$, it is expected~\cite{4authors} that conditionally on $X_n = \lfloor xn\rfloor$, $N_n$ converges weakly to a limiting distribution; furthermore, under the Gromov--Hausdorff metric,
  conditionally on $X_n = \lfloor xn\rfloor$, $\frac1n \, \mathbb{T}^{\mathrm{open}}_n$ converges weakly to a random  tree   $\mathcal{T}$. 
(See \cite{legall-miermont}
for the formalism of the Gromov--Hausdorff metric  applied to  random trees.)

The law of this limiting tree
 $\mathcal{T}$,  which turns out to be the same as found for the continuous-time model \cite{HMP}, can be   characterized  as follows: 
\medskip

\begin{itemize}

\item The spatial value of the tree $\mathcal{T}$ at the root is $x>0$, and the height of the tree is $1$;

\item The spatial value of the tree increases linearly (with coefficient $1$) along each branch until reaching a branching place;

\item Along each branch, branching occurs at rate $\frac{2\mu_s}{(1-s)^2}$ at height $s$, where $\mu_s$ stands for the spatial value at height $s$ on the branch;

\item At each branching, which is binary, the spatial value is split into two random parts according to the uniform law; 


\item Branching places and split of the spatial values are independent of each other and of everything else.  

\end{itemize}

\noindent See Figure \ref{tree3} for an example.

\begin{figure}[h]
\centerline{\includegraphics[width=10cm]{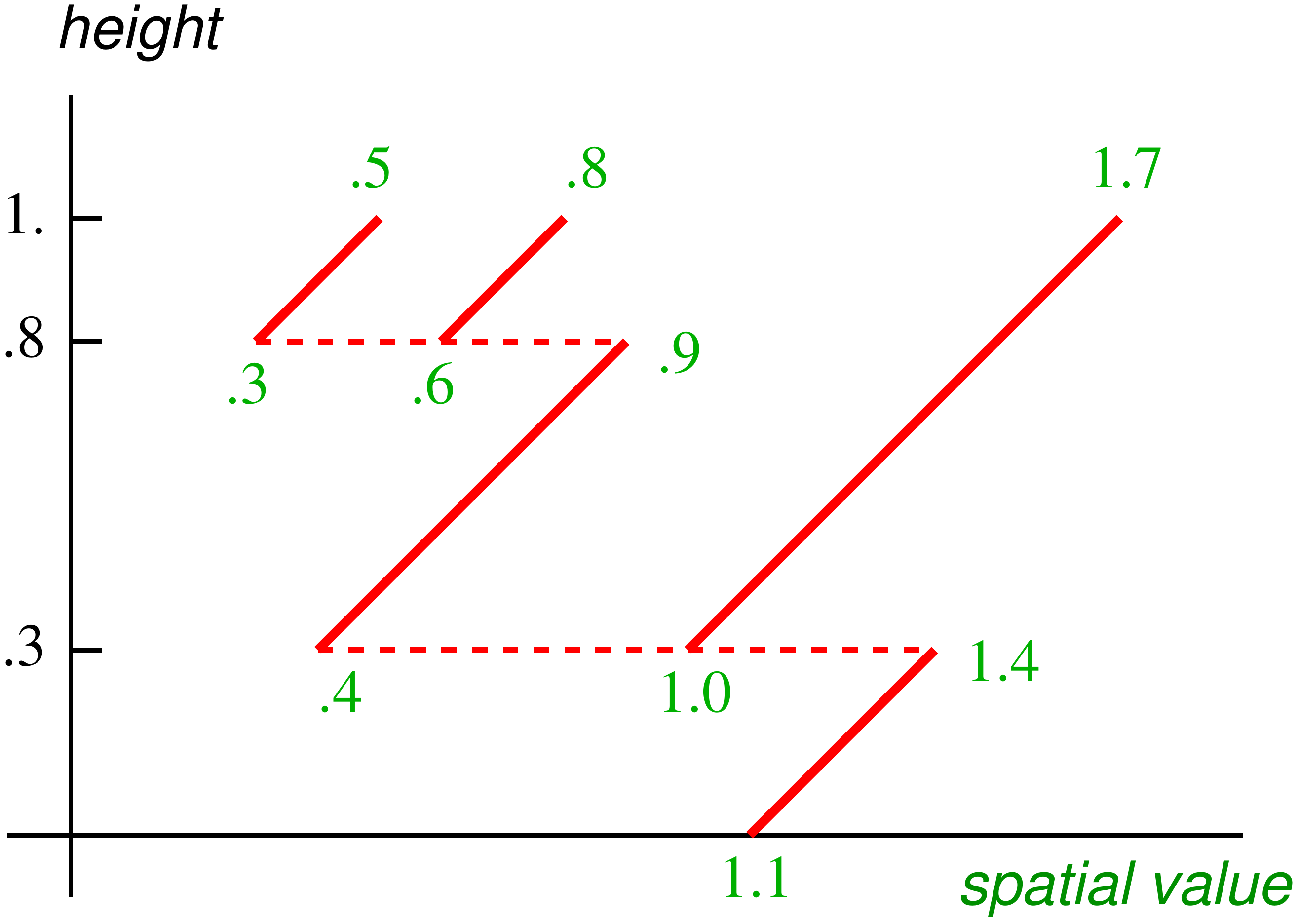}}
\caption{\small An example of $\mathcal{T}$ of height $1$ with spatial value  $1.1$ at the root. The tree, represented in red, has 3 leaves, with spatial values marked in green. Branchings happen at heights $.3$ and $.8$ .}
\label{tree3}
\end{figure}

It is also expected that upon suitable integrability conditions 
( $\langle X_0^3 \, 2^{X_0}\rangle <\infty$)
on $X_0$,  

\begin{equation}
\langle N_n\rangle \to A\in (0, \, \infty), \qquad n\to \infty,
\label{A}
\end{equation}
\noindent for some constant $A\in (0, \, \infty)$ depending on the law of $X_0$
 and that for  $\lambda = O(n^{-1/2}) $ 
\begin{equation}
\langle e^{\lambda N_n} \rangle -1 \sim -{4 \over n^2} + {3 A \lambda \over \sin^2 \left({n \sqrt{3 A \lambda}  \over 2}\right)} \ 
\label{formule}
\end{equation}
   for all integer $q\ge 2$, 
\begin{equation}
\langle N_n^q\rangle \sim b(q) \ A^q \  n^{2(q-1)}, \qquad n\to \infty,
\label{moment_Nn}
\end{equation}
where the  $b(q) \in (0, \, \infty)$ are  constants depending on $q$. 
Their expressions  can be obtained from the Taylor expansion of (\ref{formule}).

\subsection{Results}

There has been no rigorous result on the scaling limit of $\mathcal{T}$. On the other hand, it is, from a  technical point of view,  convenient to study $X_n$ biased by $N_n$ or by $N_n^{(k)}$ for a given $k\ge 0$. For example, the following recursive formula has been obtained in \cite{6authors}: for any integer $k\ge 0$,
\begin{equation}
\langle (1+X_n)2^{X_n} N_n^{(k)} \rangle = \langle (1+X_0)2^{X_0} N_0^{(k)} \rangle \prod_{i=0}^{n-1} \langle 2^{X_i} \rangle , \qquad n\ge 1.
\label{recursion_Nn}
\end{equation}
Summing over $k\ge 0$ on both sides of (\ref{recursion_Nn}), it is immediately seen that the identity holds also for $N_n$ and $N_0$ replacing $N_n^{(k)}$ and $N_0^{(k)}$, respectively. By definition, $\langle (1+X_0)2^{X_0} N_0^{(k)} \rangle = (k+1)2^k P(X_0=k)$, whereas $\prod_{i=0}^{n-1} \langle 2^{X_i} \rangle$ has order of magnitude $n^2$ for large $n$ (see (\ref{Pi}) for a more precise statement) under the additional assumption $\langle X_0^3 \, 2^{X_0} \rangle <\infty$; so the formula (\ref{recursion_Nn}) gives some information about $X_n$ when biased by $N_n^{(k)}$ or by $N_n$.

Another interesting inequality was proved in \cite{6authors2}: for $n\ge 0$ and $\ell \ge 0$,
\begin{equation}
\langle N_n^{(0)}\, {\bf 1}_{\{ X_n =\ell\} } \rangle \le \frac{1}{2^\ell} \, .
\end{equation}
In particular, we get $\langle N_n^{(0)} \rangle \le 1$ for all $n\ge 0$.

\section{Heavy tails with a transition}
\label{s:stable}

In probability theory, the central limit theorem says that sum of i.i.d.\ real-valued random variables, suitably normalized, converges weakly to the Gaussian distribution under the assumption of finite second moment, and that we get a stable distribution in the limit law if the latter assumption is conveniently weakened. This is a common phenomenon in many probability settings. For our system (\ref{recur}), the assumption $\langle X_0^3 \, 2^{X_0}\rangle <\infty$ plays the role of finite second moment in the central limit theorem  (one should imagine $\langle X_0^3 \, 2^{X_0}\rangle$ as the second moment of $X_0$ under an appropriate measure change, though nothing rigorous has been proved in this sense), under which the Derrida--Retaux conjecture for the free energy has been proved whereas a different exponent shows up under a weaker integrability assumption (see Section \ref{s:energy} for discussions on the free energy). In this section, we discuss the system under the weaker integrability assumption i.e.  $\langle X_0\, 2^{X_0} \rangle < \infty$ but  $\langle X_0^3\, 2^{X_0} \rangle =\infty$, and refer the system to the ``stable system".

Throughout this section, we assume $P(X_0=k) \sim c_0 \, 2^{-k}k^{-\alpha}$, $k\to \infty$, for some $2<\alpha \le 4$ and a constant $c_0 \in (0, \, \infty)$. Therefore (\ref{cond2}) is satisfied  but not condition (\ref{cond3}) (as  $\langle X_0^3 \, 2^{X_0}\rangle=\infty$).}  

\subsection{Conjectures}

It has been mentioned in 
  (\ref{factor_alpha}) that
$$
P(X_n>0) \sim \frac{c(\alpha)}{n^2} , \qquad n\to \infty ,
$$
with $c(\alpha) := \frac{\alpha(\alpha-2)}{2}$. 

More generally, one expects \cite{4authors} that the exponential function profile $e^{-2k/n}$ in (\ref{P(Xn=k)}) for $2^k \, P(X_n =k \, | \, X_n>0)$ should be replaced by a more complicated function 
  \begin{equation}
     P(X_n =k) = \frac{4}{n^2} \, \frac{1}{2^k} 
 \ F\left({k \over n}\right), \qquad n\to \infty,
 \label{P(Xn=k)bis)}
  \end{equation}
where  $F(x)$ is solution of the
\begin{equation}
 x F' + 2  F + F'
+  {1 \over 2}
\int_0^x F(y)
 F(x-y)   \ d y
=0 \ \ \ \ \ \ \ 
{ \rm with} \ \ \ \ F(0)= {\alpha (\alpha-2) \over 2} .
\label{F-eq}
\end{equation}
 (the  Laplace transform of $F$ can be expressed in terms of Bessel functions \cite{4authors}).

The number of open branches also changes in stable systems. For example, unlike in (\ref{A}) and (\ref{moment_Nn}) we believe that for the following large $n$ behaviors  of the expected number of open branches
\begin{equation}
\langle N_n\rangle \asymp  n^{\alpha-4}
\end{equation}
 and of the higher moments  for any integer $q\ge 2$, 
\begin{equation}
\langle N_n^q \rangle \asymp  n^{q (\alpha- 2)  - 2} 
\end{equation}
 where $a_n \asymp b_n$ means that $0 < \liminf{a_n\over b_n} < \limsup {a_n \over b_n} < \infty$ . 

Let $x>0$. The conditional weak convergence of $\frac1n \, \mathbb{T}^{\mathrm{open}}_n$ given $X_n = \lfloor xn\rfloor$ is still expected to hold, and the limiting tree $\mathcal{T}^{(\alpha)}$ would behave like $\mathcal{T}$ as described in Section \ref{s:open_tree} except for two aspects \cite{4authors}: for each branch, the branching rate is an inhomogeneous more complicated function  of $\mu_s$  and the split of the spatial value at each branching place is not according to the uniform law any more, but rather according to a law involving the function $F$.

\subsection{Results}

We have already mentioned in Section \ref{s:energy} (see (\ref{stable})) that the free energy in the stable system exhibits  a new exponent that is different from the one predicted in the Derrida--Retaux conjecture:
$$
{\cal F}_\infty = \exp\Big( - \frac{1}{\Delta^{\theta +o(1)}} \Big), \qquad \Delta \to 0+,
$$
with $\theta = \theta(\alpha) := \frac{1}{\alpha-2}$. 

Another known result for the stable system concerns the product of moment generating functions: there exist constants $0<c_9(\alpha) \le c_{10} (\alpha) <\infty$ depending on $\alpha$ and on the law of $X_0$ such that  \cite{nouveau-papier}
\begin{equation}
c_9(\alpha)\, n^{\alpha-2} \le \prod_{i=0}^{n-1}\langle 2^{X_i}\rangle \le c_{10} (\alpha) \, n^{\alpha-2}\, , \qquad n\ge 1\, .
\end{equation}
This is to be compared with the formula (\ref{Pi}), proved under the ``assumption of finite second moment" $\langle X_0^3 \, 2^{X_0}\rangle<\infty$.

\section{Heavy tail with no transition}
\label{s:heavytail}

So far, we have  always supposed $\langle X_0 \, 2^{X_0}\rangle <\infty$. In this section, we consider the case where $\langle X_0 \, 2^{X_0}\rangle =\infty$. It is convenient to formulate the discussions in a parametric form. Let $X_0^*$ be a random variable taking values in $\{ 1, \, 2, \, \ldots\}$ such that $\langle X_0^* \, 2^{X_0^*}\rangle =\infty$ , and let $p\in [0, \, 1]$. Suppose the law of $X_0$ is given by
$$
P_{X_0} = p \, P_{X_0^*} + (1-p) \delta_0,
$$
where for each random variable $\xi$, $P_\xi$ denotes its law, and $\delta_0$ is the Dirac measure. The free energy ${\cal F}_\infty$ being a non-decreasing function of $p$, there exists $p_c \in [0, \, 1]$ such that ${\cal F}_\infty=0$ for $p<p_c$ and ${\cal F}_\infty >0$ for $p>p_c$.

As $\langle X_0^* \, 2^{X_0^*}\rangle =\infty$, 
being on 
the  critical manifold $\Delta=0$ in (\ref{Delta-def},\ref{manifold}) implies that $p_c=0$ (otherwise  
$\langle X_0 \, 2^{X_0}\rangle $ would be infinite and $\Delta$ would also be infinite).

When $p_c=0$,  a new exponent appears in the slightly supercritical system~\cite{yz_bnyz}
 assuming $c_{11} k^{-\beta} 2^{-k} \le P(X_0^* \ge k) \le c_{12} k^{-\beta} 2^{-k}$ for some $0<c_{11} \le c_{12} <\infty$, $\beta\in (-\infty, \, 2)$ and all sufficiently large $k$,
\begin{equation}
{\cal F}_\infty = \exp\Big( - \frac{1}{p^{\theta_1 +o(1)}} \Big), \qquad p \to 0+,
\end{equation}
with $\theta_1 = \theta_1(\beta) := \frac{1}{2-\beta}$.

\section{Further discussions and questions}
\label{s:autres}

The recursion formula (\ref{recur}) defining our system $X_{n+1} = \max [X_{n-1}^{(1) }+ X_{n-1}^{(2)} -1, \, 0]$, in the setting of the hierarchical representation (Section \ref{s:open_tree}), has the natural interpretation that each individual inherits the total of the parents's wealth  while paying a unit amount of tax, unless  the parents's wealth is null in which case the individual pays no tax.

\subsection{Paying more tax}

What happens to a system defined by $X_{n+1} = \max [X_{n-1}^{(1) }+ X_{n-1}^{(2)} -2, \, 0]$? If the law of $X_0$ is supported in $\{ 0, \, 2, \, 4, \, \ldots\}$, then by considering $Y_n := \frac{X_n}{2}$, we recover our original system. However, if the law of $X_0$ is not supported in $\{ 0, \, 2, \, 4, \, \ldots\}$, it is even not clear how to characterize the critical manifold.
More generally what happens if $X_0$ is real rather than integer?

\subsection{Many parents}

The recursion formula defining our system $X_{n+1} = \max [ X_{n-1}^{(1) }+ X_{n-1}^{(2)} -1, \, 0]$, can obviously be extended~\cite{collet-eckmann-glaser-martin} to $X_{n+1} = \max [ X_{n-1}^{(1) }+\cdots + X_{n-1}^{(m)} -1, \, 0]$, where $m\ge 2$ is a fixed integer. The free energy in (\ref{free-energy}) becomes accordingly ${\cal F}_\infty \equiv \lim_{n \to \infty} \frac{\langle X_n\rangle}{m^n}$. All the results and predictions mentioned in this paper can be formulated for $m$ in place of $2$: some of these replacements are straightforward, while others require new ideas. For example, the critical manifold in (\ref{manifold}) is now given by $m(m-1) H'_0(m) - H_0(m)=0$.

We mention that when $m\ge 3$, the analogue of the conjectured scaling limiting open subtree $\mathcal{T}$, introduced in Section \ref{s:open_tree}, should still be binary.

\subsection{Random number of parents}

In the many parents model, if $m \ge 1$ (or even: $m\ge 0$) is a random variable independent of $(X_i^{(m)}, \, i\ge 1)$, then we get a system whose genealogical tree is a Galton--Watson tree. The free energy is ${\cal F}_\infty \equiv \lim_{n \to \infty} \frac{\langle X_n\rangle}{\langle m\rangle^n}$. It is possible to get non trivial necessary or sufficient conditions for the critical manifold, but  its precise characterization is not yet known.

\subsection{General hierarchical models}

Concerning the original model~\cite{derrida-hakim-vannimenus} $X_{n+1} = G(X_n^{(1)} + X_n^{(2)})$ with $G(\cdot)$ given in (\ref{G}) for example, all questions remain open (characterization of the critical manifold,  free energy, open subtree, universality classes, etc).


\section{Appendix}
\begin{figure}[h]
\centerline{\includegraphics[width=10.5cm]{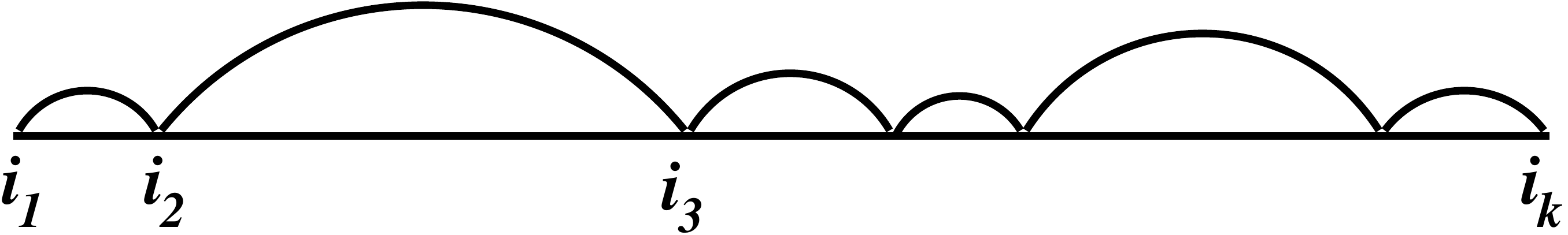}}
 \caption{\small A configuration of the molecule is specified by the number $k$ of contacts and the positions $i_1, \cdots i_k$ of these contacts.}
\label{DNA}
\end{figure}

In this appendix we present a short review of the depinning problem in presence of disorder and we 
explain how it is related to the recursion (\ref{recur}) on which this paper is focused.
In presence of disorder, the Poland Scheraga model \cite{ps}, which is a simple model of depinning or of the denaturation of the DNA molecule,  can be formulated as follows: there is a random  energy $\epsilon_i$   on  each site $i$ of a one dimensional lattice of $L$ sites. These energies are i.i.d. random variables $\epsilon_1, \epsilon_2,\cdots \epsilon_L$. Then a long molecule of length $L$ can touch this one dimensional lattice at $k$ points $i_1,i_2 \cdots i_k$ (see Figure \ref{DNA}). A configuration of the molecule is specified by the number $k$  and   the positions 
$i_1,i_2 \cdots i_k$ of the contacts and its weight is given by
$$W_k(i_1, \cdots i_k) = \exp\left[ -\beta \sum_{n=1}^k \epsilon_{i_n}\right]  \ \prod_{n=1}^{k-1} \omega(i_{n+1}-i_n)$$
where the energies appear in the exponential term while   the   product  represents the entropy factors of the loops between successive contacts.

A typical choice for these entropy factors is
\begin{equation}
\omega(i)= i^{-c}
\label{omega}
\end{equation}
for some $c >1$
(in fact it is only the large $i$  decay which has an influence on the nature of the transition).
Then  the partition   function $Z_L$ is given by
$$Z_L (\epsilon_1, \cdots \epsilon_L) \equiv  \sum_{k\ge 2} \ \sum_{1=i_1 < i_2 < \cdots < i_k=L} W_k(i_1,\cdots,i_k)$$
(here all configurations have a contact at $i_1=1$ and at $i_k=L$).
As the inverse temperature $\beta $  or the distribution of the energies $\epsilon_i$ are changed, the system undergoes a phase transition between a phase where   ${\cal F}_\infty=0 $
 and a phase where ${\cal F}_\infty >0$. Here the free energy ${\cal F}_\infty$ is defined by
$${\cal F}_\infty \equiv \lim_{L \to \infty} {\langle \log Z_L \rangle \over L} $$
where the average is over the $\epsilon_i$.

When all the $\epsilon_i$ are equal, i.e. in absence of disorder (this is called the pure case), the   free  energy ${\cal F}_\infty $ can be calculated exactly \cite{ps}, the critical manifold can be fully characterized and the system exhibits a first order phase transition  or a second order phase transition depending on the value of $c$.

In presence of disorder, the two main questions  are:
\begin{enumerate}
\item  How to characterize the critical manifold? In particular one would like to know  how the critical manifold is shifted  in the case of a weak disorder, i.e. when the distribution of the $\epsilon_i$'s is narrow.
\item What is the nature of the transition  in presence of disorder? In particular  does a weak disorder change the nature of the transition?
\end{enumerate}
A number of predictions have been made in the physics literature, sometimes contradictory,  \cite{luck,derrida-hakim-vannimenus,IM,monthus,MG1,KL,tang-chate}. It is however  now well established under which condition the nature of the transition is the same for the pure system and in the case of a weak disorder \cite{giacomin_stf,giacomin-toninelli-lacoin,G3,GLT,GLT2}
as well as how the  transition point is shifted  \cite{Alex,AZ,BL,dglt}.

Concerning the nature of the transition,  the main result \cite{giacomin-toninelli,GT1} is that in presence of disorder, the transition is always smooth, implying the impossibility of first order transitions or of diverging specific heats. Still the precise nature of the singularity is not understood, in particular  the 20-years old prediction of  a transition of  the Berezinskii--Kosterlitz--Thouless type  \cite{tang-chate} has not yet been confirmed mathematically.

All the above questions can be asked for the version of the depinning problem  on a hierarchical lattice \cite{derrida-hakim-vannimenus,MG2}. In this case, the partition function (defined only when $L$ is a power of $2$) satisfies a simple recursion
$$Z_{2 L} = {Z_L^{(1)} \, Z_L^{(2)} + b-1 \over b} $$ where $Z_L^{(1)}$ and $Z_L^{(2)}$ are two independent realizations of $Z_L$ and the $Z_1$'s are i.i.d. random variables  given by
$$Z_1= e^{-\beta \epsilon} \ . $$
Here $b$ plays a  role similar to  the parameter $c$  in the  entropy factor (\ref{omega}).
Clearly if one defines $X_n$ by
$$X_n= \log Z_{2^n}$$
it satisfies the recursion (\ref{system},\ref{G}).
As for the original problem,  the same  questions can be asked.  A number of results already exist on the shift of the transition and on the condition for the  transition to remain the same for the pure system and in presence of a weak disorder \cite{BT,giacomin-lacoin-toninelli,L,LT}. Still an understanding of the nature of the transition 
for strong disorder or even for weak disorder when disorder is relevant is lacking.



\bigskip
\bigskip

\begin{thebibliography}{99}
\label{references}

\baselineskip=14pt

\bibitem{Alex}
    Alexander, K.S.\ (2008). 
    The effect of disorder on polymer depinning transitions. 
    {\it Commun.\ Math.\ Phys.}
 {\bf 279}, 117--146.

\bibitem{AZ}
Alexander, K.S.,  Zygouras, N. (2009). Quenched and annealed critical points in polymer pinning models. 
    {\it Commun.\ Math.\ Phys.}
 {\bf 291}, 659-689.

\bibitem{BL}
    Berger, Q.\ and  Lacoin, H.\ (2018). 
    Pinning on a defect line: characterization of marginal disorder relevance and sharp asymptotics for the critical point shift. 
    {\it J.\ Inst.\ Math.\ Jussieu} {\bf 17}, 305--346.

\bibitem{BT}
Berger, Q.,  Toninelli, F. L. (2013). Hierarchical pinning model in correlated random environment. 
{\it Ann. Inst. H. Poincar\'e Probab. Statist.}
  {\bf 49},   781-816.



\bibitem{bmxyz_questions}
    Chen, X., Derrida, B., Hu, Y., Lifshits, M., and Shi, Z. (2019). 
    A max-type recursive model: some properties and open questions. 
    In: Sojourns in Probability Theory and Statistical Physics-III (pp.~166--186). Springer, Singapore.

\bibitem{6authors}
    Chen, X., Dagard, V., Derrida, B., Hu, Y., Lifshits, M.\ and Shi, Z.
    The Derrida--Retaux conjecture on recursive models.
    arXiv:1907.01601

\bibitem{6authors2}
    Chen, X., Dagard, V., Derrida, B., Hu, Y., Lifshits, M.\ and Shi, Z.
    The sustainability probability in the critical Derrida--Retaux system.
    In preparation

\bibitem{4authors}
    Chen, X., Dagard, V., Derrida, B.\ and Shi, Z. (2020)
    The critical behaviors and the scaling functions of a coalescence equation.
    arXiv:2001.00853, to appear in  J. Phys. A: Math. Theor.

\bibitem{nouveau-papier}
    Chen, X.,  and Shi, Z.
The stable Derrida--Retaux conjecture. In preparation

\bibitem{collet-eckmann-glaser-martin2}
    Collet, P., Eckmann, J.P., Glaser, V.\ and Martin, A.\ (1984).
    A spin glass with random couplings.
    {\it J.\ Statist.\ Phys.} {\bf 36}, 89--106.

\bibitem{collet-eckmann-glaser-martin}
    Collet, P., Eckmann, J.P., Glaser, V.\ and Martin, A.\ (1984).
    Study of the iterations of a mapping associated to a spin-glass model.
    {\it Commun.\ Math.\ Phys.} {\bf 94}, 353--370.

\bibitem{dglt}
    Derrida, B., Giacomin, G., Lacoin, H.\ and Toninelli, F.L.\ (2009).
    Fractional moment bounds and disorder relevance for pinning models. 
    {\it Commun.\ Math.\ Phys.} {\bf 287}, 867--887.

\bibitem{derrida-hakim-vannimenus}
    Derrida, B., Hakim, V.\ and Vannimenus, J.\ (1992).\
    Effect of disorder on two-dimensional wetting.
    {\it J.\ Statist.\ Phys.} {\bf 66}, 1189--1213.

\bibitem{derrida-retaux}
    Derrida, B.\ and Retaux, M.\ (2014).\
    The depinning transition in presence of disorder: a toy model.
    {\it J.\ Statist.\ Phys.} {\bf 156}, 268--290.

\bibitem{luck}
    Forgacs, G., Luck, J.M., Nieuwenhuizen, T.M.\ and Orland, H.\ (1986).  
    Wetting of a disordered substrate: exact critical behavior in two dimensions. 
    {\it Phys.\ Rev.\ Lett.}, {\bf 57}, 2184.

\bibitem{giacomin}
    Giacomin, G.\ (2007). 
    {\it Random Polymer Models.}
    Imperial College Press.

\bibitem{giacomin_stf}
    Giacomin, G.\ (2011).
    {\it Disorder and critical phenomena through basic probability models.}
    {\it \'Ecole d'\'et\'e Saint-Flour XL (2010)},
    Lecture Notes in Mathematics {\bf 2025}, 
    Springer, Heidelberg.  

\bibitem{giacomin-toninelli}
    Giacomin, G.\ and Toninelli, F.L.\ (2006). 
    Smoothing effect of quenched disorder on polymer depinning transitions.
    {\it Commun.\ Math.\ Phys.} {\bf 266}, 1--16.

\bibitem{GT1}
Giacomin, G.,  Toninelli, F. L. (2006). Smoothing of depinning transitions for directed polymers with quenched disorder. {\it Phys. Rev. Lett.} , {\bf 96}, 070602.

\bibitem{giacomin-toninelli-lacoin}
    Giacomin, G., Toninelli, F.\ and Lacoin, H.\ (2010). 
    Marginal relevance of disorder for pinning models. 
    {\it Commun.\ Pure Appl.\ Math.}, {\bf 63}, 233--265.

\bibitem{GLT}
Giacomin, G., Lacoin, H.,  Toninelli, F. L. (2011). Disorder relevance at marginality and critical point shift. 
{\it Ann. Inst. H. Poincar\'e Probab. Statist.}
 {\bf  47},   148-175.

\bibitem{GLT2}
Giacomin, G.,  Toninelli, F. L. (2009). On the irrelevant disorder regime of pinning models. {\it Ann.  Probab.}, {\bf 37}, 1841-1875.

\bibitem{G3}
Giacomin, G. (2009). Renewal sequences, disordered potentials, and pinning phenomena. In Spin Glasses: Statics and Dynamics (pp. 235-270). Birkh\"auser Basel.


 \bibitem{giacomin-lacoin-toninelli}
     Giacomin, G., Lacoin, H.\ and Toninelli, F.L.\ (2010).\
     Hierarchical pinning models, quadratic maps and quenched disorder.
     {\it Probab.\ Theory Related Fields}
 {\bf 147}, 185--216.

\bibitem{HMP}
    Hu, Y., Mallein, B.\ and Pain, M.\ (2018+).
    An exactly solvable continuous-time Derrida--Retaux model.
    {\tt arXiv:1811.08749}

\bibitem{yz_bnyz}
    Hu, Y.\ and Shi, Z.\ (2018).\
    The free energy in the Derrida--Retaux recursive model.
    {\it J.\ Statist.\ Phys.} {\bf 172}, 718--741.

\bibitem{IM}
Igl\'oi, F.,  Monthus, C. (2005). Strong disorder RG approach of random systems. {\it Phys. Rep.}, 
{\bf 412}, 277-431.

\bibitem{KL}
Kunz, H.,  Livi, R. (2012). DNA denaturation and wetting in the presence of disorder.  {\it Eur. Phys. Lett.}, {\bf 99}, 30001.


\bibitem{legall-miermont}
    Le Gall, J.-F.\ and Miermont, G.\ (2012).
    Scaling limits of random trees and planar maps. 
    In: {\it Probability and Statistical Physics in Two and More Dimensions} 155--211. Clay Mathematics Proceedings {\bf 15}, Amer.\ Math.\ Soc., Providence, RI.

\bibitem{L}
Lacoin, H. (2010). Hierarchical pinning model with site disorder: disorder is marginally relevant.
     {\it Probab.\ Theory Related Fields}
 {\bf 148}, 159-175.


\bibitem{LT}
Lacoin, H.,  Toninelli, F. L. (2009). A smoothing inequality for hierarchical pinning models. In Spin glasses: statics and dynamics (pp. 271-278). Birkh\"auser Basel.

\bibitem{monthus}
    Monthus, C.\ (2017). 
    Strong disorder renewal approach to DNA denaturation and wetting: typical and large deviation properties of the free energy.
    {\it J.\ Statist.\ Mech.\ Theory Exper.} 2017, 013301.

\bibitem{MG1}
Monthus, C.,  Garel, T. (2005). Distribution of pseudo-critical temperatures and lack of self-averaging in disordered Poland-Scheraga models with different loop exponents. {\it  Eur. Phys. J. B} {\bf 48}, 393-403.

\bibitem{MG2}
Monthus, C.,  Garel, T. (2008). Critical points of quadratic renormalizations of random variables and phase transitions of disordered polymer models on diamond lattices. {\it Phys. Rev. E} {\bf 77}, 021132.

\bibitem{ps}
  Poland D.,  Scheraga H.A. (1970). Theory of helix-coil transitions in biopolymers; Statistical mechanical
theory of order-disorder transitions in biological macromolecules, Academic Press, 1970

\bibitem{tang-chate}
    Tang, L.H.\ and Chat\'e, H.\ (2001). 
    Rare-event induced binding transition of heteropolymers. 
    {\it Phys.\ Rev.\ Lett. } {\bf 86}, 830.


\end{thebibliography}
\end{document}